\documentclass[hidelinks,11pt,a4paper,oneside]{article}

\pdfoutput=1
\usepackage{amssymb}
\usepackage{amsmath}
\usepackage{cite}
\usepackage{graphicx}
\usepackage{braket}
\usepackage{amsbsy}
\usepackage{bm} 
\usepackage{euscript} 
\usepackage[a4paper,top=1in,bottom=1.25in,left=1in,right=1in]{geometry}
\usepackage[perpage,para]{footmisc}
\usepackage{lipsum}
\usepackage[margin=0.5in,font=small,labelfont=bf]{caption}
\usepackage{enumitem}

\usepackage[utf8]{inputenc}
\usepackage{dcolumn}
\usepackage{array}
\usepackage{multirow}

\newcommand*{\DiffOp}{\mathop{\mathrm{d{}}}\nolimits}
\newcommand*{\diff}[1]{\DiffOp\mathclose{#1}}
\newcommand*{\deriv}[3][]{\frac{\DiffOp}}

\begin{document}

\title{\textbf{Nudged elastic band calculations accelerated with Gaussian process regression}}

\author{Olli-Pekka Koistinen $^{1,2,3}$ \and Freyja~B. Dagbjartsd\'ottir $^2$ \and Vilhj\'almur \'Asgeirsson $^2$ \and Aki Vehtari $^1$ \and Hannes J\'onsson $^{2,3}$}

\date{}

\maketitle

\begin{center}
$^1$ Helsinki Institute for Information Technology HIIT,\\ Department of Computer Science, Aalto University, Espoo, Finland
\end{center}
\begin{center}
$^2$ Science Institute and Faculty of Physical Sciences, University of Iceland, Reykjav\'{\i}k, Iceland
\end{center}
\begin{center}
$^3$ Department of Applied Physics, Aalto University, Espoo, Finland
\end{center}
\begin{center}
hj@hi.is
\end{center}

\hfill

\begin{abstract}
Minimum energy paths for transitions such as atomic and/or spin rearrangements in thermalized systems are the transition paths of largest statistical weight.
Such paths are frequently calculated using the nudged elastic band method, where an initial path is iteratively shifted to the nearest minimum energy path.
The computational effort can be large, especially when {\it ab initio} or electron density functional calculations are used to evaluate the energy and atomic forces.
Here, we show how the number of such evaluations can be reduced by an order of magnitude using
a Gaussian process regression approach where an approximate energy surface is generated and refined in each iteration.
When the goal is to evaluate the transition rate within harmonic transition state theory,
the evaluation of the Hessian matrix at the initial and final state minima can be carried out
beforehand and used as input in the minimum energy path calculation, thereby improving stability
and reducing the number of iterations needed for convergence. A Gaussian process model also provides an uncertainty
estimate for the approximate energy surface, and this can be used to focus the calculations on the
lesser-known part of the path, thereby reducing the number of needed energy and force evaluations to
a half in the present calculations.
The methodology is illustrated using the two-dimensional M\"uller-Brown potential surface and performance
assessed on an established benchmark involving 13 rearrangement transitions of a heptamer island on a solid surface.
\end{abstract}

\hfill

\section{Introduction}
\label{sec:Introduction} 

Theoretical studies of the transition mechanism and estimation of the rate of thermally activated events involving
displacements of atoms or rotations of magnetic moments 
often involve finding a minimum energy path (MEP) connecting initial and final state
minima on the energy surface characterizing the system.
An MEP is a natural choice for a reaction coordinate since it represents a path of maximal statistical weight in a system
in thermal equilibrium with a heat bath.
Transition state theory \cite{TST,Kramers40,Keck67}
calculations can be carried out using this reaction coordinate to parametrize, for example, a
hyperplanar representation of the transition state \cite{Schenter94,Mills95}.
Even though such a reaction coordinate represents only one particular mechanism for the transition, it is possible to 
discover a new mechanism corresponding to a lower free energy barrier when full variational optimization of
both the location and orientation of the hyperplanar transition state is carried out \cite{Johannesson00,Bligaard04}.
In such a case, an MEP is just a convenient tool for the implementation of a full free energy calculation.

Most often, transition rates are, however, estimated from the harmonic approximation to transition state theory,
where the maximum rise in the energy along an MEP gives the activation energy of the transition and the
pre-exponential factor in the Arrhenius expression for the rate can be obtained from the Hessian matrix evaluated at the
initial state minimum and the energy maximum -- a first-order saddle point on the energy surface \cite{HTST}.
While it is possible to use various methods to converge directly on a saddle point starting from some initial guess,
knowledge of the whole MEP is useful because it is important to make sure that the highest first-order
saddle point for the full transition has been found.  
Furthermore, calculations of MEPs often reveal unknown intermediate minima and unexpected transition mechanisms \cite{Jonsson11}
and therefore play an important role in the studies of the mechanism and rate of thermally activated
transitions.
Most often, such calculations are carried out for transitions involving rearrangements of atoms, 
but similar considerations apply to thermally activated transitions
where magnetic moments rotate
from one magnetic state to another \cite{bessarab_12,bessarab_13a,bessarab_14,bessarab_14b}.

The nudged elastic band (NEB) method is a commonly used iterative approach to find MEPs \cite{Jonsson11,Mills95,NEBleri}.
For magnetic transitions, a variant of the method called geodesic NEB has been developed \cite{Bessarab15}.
In the NEB method, the path between two local minima on the energy surface is represented by
a discrete set of replicas of the system, referred to as `images', each of them consisting of values for
all degrees of freedom.
Starting from some initial path, the locations of the images on the energy surface are iteratively optimized
so as to obtain a discrete representation of an MEP.

Each NEB calculation typically involves on the order of a hundred evaluations of the energy and force
(the negative gradient of the energy)
for each one of the images, and the path is typically represented by five to ten images.
The evaluations were initially performed mostly using analytical potential energy functions,
but nowadays electronic structure calculations are also used extensively in NEB applications.
Since a typical electronic structure calculation takes on the order of tens of CPU minutes or more,
the NEB calculations can become computationally demanding.
In addition, the calculation may need to be repeated if there are several possible final states for the transition.
Thus, it would be valuable to find ways to accelerate NEB calculations.
To get the most out of the computationally intensive electronic structure calculations,
the information obtained from them should be exploited better to decrease the number of NEB iterations
instead of forgetting it after one iteration.

The use of machine learning to accelerate MEP
and saddle point calculations has been introduced by Peterson \cite{Peterson16}, 
who applied neural networks to construct an approximate energy surface
for which NEB calculations were carried out.
After relaxation of the path on the approximate energy surface,
the true energy and force were evaluated at the images of the relaxed path
to see whether or not the path had converged on an MEP on the true energy surface.
If true convergence had not been reached, the new energy and force values calculated at the location of the images
were added to the training data set and the model was updated.
This procedure was repeated iteratively until the approximate energy surface
was accurate enough for converging on the true MEP.

Proof-of-principle results have also been presented where
Gaussian process regression (GPR) \cite{OHagan78,MacKay98,Neal99,Rasmussen06} 
is applied to accelerate NEB calculations \cite{Koistinen16}.
Since the calculations are largely based on the gradient vector of the
energy surface, straightforward inclusion of derivative observations and
prediction of derivatives can be seen as advantages of GPR for this application.
It is also easy to encode prior assumptions about the smoothness properties of
the energy surface into the covariance function of the Gaussian process (GP) model
or learn about these properties from the data.
Analytical expressions for the posterior predictions conditional on the hyperparameters of the GP model
allow both fast predictions and reliable estimation of uncertainties.
The predictive performance of GPR has been shown to be competitive with other machine learning methods
especially when the number of observations is small \cite{Lampinen01}.

The GPR approach to MEP calculations is extended here
by presenting two algorithms to accelerate climbing image nudged elastic
band calculations (CI-NEB), where one of the images is made to converge to a small tolerance on the highest energy
maximum along the MEP \cite{ClimbingImage}.
The basic GPR approach is described as the all-images-evaluated (AIE)
algorithm, where the energy and force are evaluated at all intermediate
images of the CI-NEB
before the approximation to the energy surface is updated. 
In a more advanced algorithm,
the energy and force are evaluated at only one image before a new
approximate energy surface is constructed.
We refer to the latter as the one-image-evaluated (OIE) algorithm.
As a probabilistic model, a GP expresses the energy predictions
as probability distributions, which means that the uncertainty
of the prediction can also be estimated, e.g., as the variance of
the posterior distribution. This uncertainty estimate is used by the
OIE algorithm to select the image to be evaluated in such a way as to give maximal
improvement of the model.
By directing the evaluations to locations where they are most needed,
the OIE algorithm skips some of the energy and force evaluations
and thus decreases the overall computation time compared to the
AIE algorithm.
This approach has similarities with Bayesian optimization \cite{Shahriari16},
where the uncertainties of a GP model are used to define an acquisition function
that is used to select the locations of new evaluations
in a global optimization task.

Another extension of the GPR approach presented here applies when the overall goal is to estimate the 
forward and backward transition rates using harmonic transition state theory.
Then, the Hessian matrix needs to be evaluated at the initial and final state minima, as well as at the highest first-order 
saddle point along the MEP. The evaluation of the Hessian at the endpoint minima can be carried out before the MEP calculation
to provide additional input information into the GP model about the shape of the energy surface in the vicinity of the two ends of the MEP.

The article is organized as follows: In Sec.~\ref{sec:NEB}, a brief introduction to the NEB method is given,
followed by presentation of necessary GP theory for the GPR approach in Sec.~\ref{sec:GPR}.
In Sec.~\ref{sec:Algorithms}, the two implementations, the AIE algorithm and the
OIE algorithm, are described and illustrated on the two-dimensional M\"uller-Brown energy surface \cite{MullerBrown79}.
In Sec.~\ref{sec:Application}, the heptamer island benchmark is described and performance statistics are given as a function of the number of
degrees of freedom.
The article concludes with a discussion in Sec.~\ref{sec:Discussion}.


\section{Nudged elastic band method}
\label{sec:NEB}

The objective of the nudged elastic band (NEB) method is to find a minimum energy path (MEP) connecting two given local minima on an energy surface.
An MEP is defined as a path for which the gradient of the energy has zero component perpendicular to the path tangent.
In the NEB method, the path is represented in a discretized way as a set of images, which are sets of values of all degrees of freedom in the system
(atom coordinates and angles specifying orientation of magnetic vectors, and possibly also simulation box size and shape).
The MEP is found iteratively, starting from some initial path between the two minima.
Most often, a straight line interpolation between the minima has been used to generate the initial path \cite{NEBleri}, but
a better approach is to start with a path that interpolates as closely as possible the distances between neighboring atoms,
the so-called image dependent pair potential (IDPP) method \cite{Smidstrup14}.

The key feature of the NEB algorithm is the `nudging', a projection which is used
to separate the force components perpendicular and parallel to the path from each other.
If each image is just moved along the force vector (negative gradient of the energy),
they would end up sliding down to the nearest minima.
The main idea in the NEB method is to take into account only the force component perpendicular
to the path and at the same time control the distribution of the images along the path.
The projection of the force requires an estimate of the local tangent to the path at the location of each image.
A well-behaved estimate is obtained by defining the tangent based on the vector to the neighboring image of higher energy
or, if both of the neighbors are either higher or lower in energy than the current image, a weighted average of the vectors to the two neighboring images \cite{ImprovedTangent}.

To control the distribution of the images along the path, a spring force acting in the direction of the path tangent is typically introduced.
The most common choice is to strive for an even distribution, but one can also choose to have, for example, a higher density of images where the energy is larger \cite{ClimbingImage}.
The spring force also prevents the path from becoming arbitrarily long in regions of little or no force.
This is important, for example, in calculations of adsorption and desorption of molecules at surfaces.

In each iteration, the images are moved along the resultant vector of the spring force
and the component of the true force perpendicular to the path,
which is here referred to as the NEB force $F_\mathrm{NEB}$.
The true force is the negative gradient of the energy, and in most cases an evaluation of the energy
also delivers the gradient vector at little or no additional expense.
It is, however, typically too expensive to evaluate the second derivatives of the energy,
and the iterative algorithms are therefore based solely on the gradient and energy at each image
(in addition to the spring force which depends on the distribution of the images).
A simple and stable method that has been used extensively in NEB calculations will be used here to control the step size of the movements.
It is based on a velocity Verlet dynamics algorithm where only the component of the velocity in the direction
of the NEB force is included as long as the inner product with the NEB force is positive \cite{NEBleri}.
A somewhat higher efficiency can be obtained by using quadratically convergent algorithms such as conjugate gradient or quasi-Newton \cite{Sheppard08},
but those can be less stable especially in the beginning of an NEB calculation.

The most important part of an MEP is the vicinity of the highest energy saddle point, especially in harmonic transition state
theory calculations where the highest energy saddle point directly gives an estimate of the activation energy of the transition.
It is, therefore, advantageous to let the highest energy image 
move to the maximum energy along the path.
This variant of the NEB method is referred to as the
climbing image nudged elastic band (CI-NEB) method \cite{ClimbingImage}.
Whereas the component of the true force acting in the direction of the tangent is normally removed from the NEB force,
for the climbing image it is instead flipped around to point towards the direction of higher energy along the path.
In the CI-NEB method, the spring force is not applied to the climbing image and the rest of the images are distributed evenly
on each side of the climbing image.
To keep the intervals reasonably similar on both sides of the climbing image, the regular NEB method is typically conducted
first (to some preliminary tolerance) so that the image selected as the climbing image is not too far from the true saddle point.
The rest of the MEP is mainly needed to 
ensure that the highest saddle point has been identified and to provide an estimate of the tangent to the path in order to
carry out the nudging projections of the forces. 
It is more important to make the climbing image converge rigorously than the other images. It is, therefore, practical to apply
a tighter tolerance for the magnitude of the NEB force acting on the climbing image than to the other images in CI-NEB calculations.

In the heptamer island benchmark presented here, the path was represented by seven images, $N_\mathrm{im}=7$, and the initial path
was generated using the IDPP method.
All spring constants were chosen to be 1~eV/{\AA} to give an
even distribution of the images along the path on each side of the climbing image.
The focus here is on calculations where the energy and force are obtained using some {\it ab initio}
or density functional theory calculations. The computational effort in all other parts of the calculation is then insignificant in
comparison, and thus the overall computational effort is well characterized by simply the number of times the energy and force need
to be evaluated in order to converge on the MEP.
Below, we describe various strategies to accelerate CI-NEB calculations with Gaussian process regression.


\section{Gaussian process regression}
\label{sec:GPR}

A Gaussian process (GP) is a flexible probabilistic model for functions in a continuous domain \cite{OHagan78,MacKay98,Neal99,Rasmussen06}.
It is defined by a mean function $m(\mathbf{x})$ and a covariance function $k(\mathbf{x}^{(i)},\mathbf{x}^{(j)})$,
so that the joint probability distribution of the function values
$\mathbf{f} = [f(\mathbf{x}^{(1)}), f(\mathbf{x}^{(2)}), \ldots , f(\mathbf{x}^{(N)})]^\mathsf{T}$
at any finite set of input points
$\mathbf{X} = [\mathbf{x}^{(1)}, \mathbf{x}^{(2)}, \ldots , \mathbf{x}^{(N)}]^\mathsf{T} \in \mathbb{R}^{N \times D}$
is a multivariate Gaussian distribution $p(\mathbf{f}) = \mathcal{N}(\mathbf{m},K(\mathbf{X},\mathbf{X}))$, where
$\mathbf{m} = [m(\mathbf{x}^{(1)}),m(\mathbf{x}^{(2)}),\ldots,m(\mathbf{x}^{(N)})]^\mathsf{T}$ and the notation $K(\mathbf{X},\mathbf{X}')$ represents
a covariance matrix with entries $K_{ij} = k(\mathbf{x}_i,\mathbf{x}_j')$.
Thus, a GP can be seen as an infinite-dimensional generalization of the multivariate Gaussian
distribution, serving as a prior probability distribution for the unknown function $f$.
After evaluating the function at some training data points,
the probability model is updated and a posterior probability distribution
can be calculated for the function value at any point.

The most important part of the GP model is the covariance function, which defines how
the function values at any two input points depend on each other, usually based on the distance between the points.
Through selection of the covariance function, different prior assumptions about the properties of the function
can be encoded into the model. To favor smooth functions, the infinitely differentiable squared exponential
covariance function
\begin{equation*}
k(\mathbf{x}^{(i)},\mathbf{x}^{(j)}) = \sigma_{\mathrm{c}}^2 + \sigma_{\mathrm{m}}^2 \exp\left(-\frac{1}{2}\sum_{d=1}^{D} \frac{(x_d^{(i)}-x_d^{(j)})^2}{l_d^2}\right)
\end{equation*}
is used here.
The hyperparameters $\bm{l} = \{ l_1, \ldots, l_D \}$ are length scales that define the range of the covariance in each dimension,
and $\sigma_{\mathrm{m}}^2$ is a hyperparameter that controls the magnitude of the covariation.
The mean function is here set to zero, but the additional constant term $\sigma_{\mathrm{c}}^2$ in the covariance function has a similar effect
as integration over an unknown constant intercept term having a Gaussian prior distribution with variance $\sigma_{\mathrm{c}}^2$.

Consider a regression problem $y = f(\mathbf{x}) + \epsilon$, where $\epsilon$ is a Gaussian noise term with variance $\sigma^2$,
and a training data set $\{\mathbf{X},\mathbf{y}\}$, where $\mathbf{y} = [y^{(1)}, y^{(2)}, \ldots, y^{(N)}]^\mathsf{T}$ includes $N$ noisy
output observations from the $N$ input points
$\mathbf{X} = [\mathbf{x}^{(1)}, \mathbf{x}^{(2)}, \ldots , \mathbf{x}^{(N)}]^\mathsf{T} \in \mathbb{R}^{N \times D}$.
The posterior predictive distribution for the function value $f(\mathbf{x}^*)$ at a new point $\mathbf{x}^*$,
conditional on the GP model hyperparameters $\bm{\theta} = \{\sigma_{\mathrm{m}}^2,\bm{l}\}$,
is a Gaussian distribution with mean
\begin{equation*}
  {\mathrm{E}}[f(\mathbf{x}^*)|\mathbf{y},\mathbf{X},\bm{\theta}] = K(\mathbf{x}^*,\mathbf{X})\left(K(\mathbf{X},\mathbf{X}) + \sigma^2 \mathbf{I}\right)^{-1}\mathbf{y}
\end{equation*}
and variance
\begin{equation*}
  \mathrm{Var}[f(\mathbf{x}^*)|\mathbf{y},\mathbf{X},\bm{\theta}] =
  k(\mathbf{x}^*,\mathbf{x}^*)-K(\mathbf{x}^*,\mathbf{X}) \left(K(\mathbf{X},\mathbf{X}) + \sigma^2 \mathbf{I}\right)^{-1}K(\mathbf{X},\mathbf{x}^*),
\end{equation*}
where $\mathbf{I}$ is the identity matrix. Here the hyperparameter values $\bm{\theta} = \{\sigma_{\mathrm{m}}^2,\bm{l}\}$
are optimized by defining a prior probability distribution $p(\bm{\theta})$ and maximizing the marginal posterior probability density
$p(\bm{\theta}|\mathbf{y},\mathbf{X}) \propto p(\bm{\theta})p(\mathbf{y}|\mathbf{X},\bm{\theta})$, where
$p(\mathbf{y}|\mathbf{X},\bm{\theta}) = \int_{\mathbf{f}} p(\mathbf{y} | \mathbf{f}) p(\mathbf{f} | \mathbf{X}, \bm{\theta}) \diff{\mathbf{f}}$
is the marginal likelihood of $\bm{\theta}$ in the light of the observed data set $\{\mathbf{X},\mathbf{y}\}$.

Since differentiation is a linear operation, the derivative of a GP is as well a GP \cite{OHagan92,Rasmussen03,Solak03,Riihimaki10,Bartok15}.
This makes it straightforward to use derivative information and predict derivatives of the function $f$.
Derivative observations can be included in the model by extending the observation vector $\mathbf{y}$
to include partial derivative observations 
and by extending the covariance matrix $K(\mathbf{X},\mathbf{X})$ correspondingly to include covariances
between the function values and partial derivatives and covariances between the partial derivatives themselves.
In the case of the squared exponential covariance function, these entries are obtained by
\begin{align*}
\mathrm{Cov}\left[\frac{\partial f(\mathbf{x}^{(i)})}{\partial
  x^{(i)}_d},f(\mathbf{x}^{(j)})\right]=&\frac{\partial}{\partial
  x^{(i)}_d}\mathrm{Cov}\left[f(\mathbf{x}^{(i)}),f(\mathbf{x}^{(j)})\right]
  = \frac{\partial k(\mathbf{x}^{(i)},\mathbf{x}^{(j)})}{\partial x^{(i)}_d} \\
  =& - \frac{\sigma_{\mathrm{m}}^2 (x_d^{(i)}-x_d^{(j)})}{l_d^2} \exp\left(-\frac{1}{2}\sum_{g=1}^{D} \frac{(x_g^{(i)}-x_g^{(j)})^2}{l_g^2}\right)
\end{align*}
and
\begin{align*}
\mathrm{Cov}\left[\frac{\partial f(\mathbf{x}^{(i)})}{\partial
  x^{(i)}_{d_1}},\frac{\partial f(\mathbf{x}^{(j)})}{\partial x^{(j)}_{d_2}}\right]=&
  \frac{\partial^2}{\partial x^{(i)}_{d_1}\partial x^{(j)}_{d_2}}\mathrm{Cov}\left[f(\mathbf{x}^{(i)}),f(\mathbf{x}^{(j)})\right] =
  \frac{\partial^2 k(\mathbf{x}^{(i)},\mathbf{x}^{(j)})}{\partial x^{(i)}_{d_1}\partial x^{(j)}_{d_2}} \\
  =& \frac{\sigma_{\mathrm{m}}^2}{l_{d_1}^2}\left(\delta_{{d_1}{d_2}}-\frac{(x^{(i)}_{d_1}-x^{(j)}_{d_1})(x^{(i)}_{d_2}-x^{(j)}_{d_2})}{l_{d_2}^2}\right) \\
  & \times \exp\left(-\frac{1}{2}\sum_{g=1}^{D} \frac{(x_g^{(i)}-x_g^{(j)})^2}{l_g^2}\right),
\end{align*}
where $\delta_{{d_1}{d_2}}=1$ if $d_1=d_2$, and $\delta_{{d_1}{d_2}}=0$ if $d_1 \neq d_2$.
These same expressions are needed also when predicting the derivatives.
The posterior predictive distribution of the partial derivative of function $f$ with respect to dimension $d$
at a new point $\mathbf{x^*}$ is a Gaussian distribution with mean
\begin{equation*}
  {\mathrm{E}}\left[ \frac{\partial f(\mathbf{x}^*)}{\partial x^*_d} \middle| \mathbf{y},\mathbf{X},\bm{\theta} \right] =
  \frac{\partial K(\mathbf{x}^*,\mathbf{X})}{\partial x^*_d}\left(K(\mathbf{X},\mathbf{X}) + \sigma^2 \mathbf{I}\right)^{-1}\mathbf{y}
\end{equation*}
and variance
\begin{equation*}
  \mathrm{Var}\left[\frac{\partial f(\mathbf{x}^*)}{\partial x^*_d} \middle|\mathbf{y},\mathbf{X},\bm{\theta}\right] =
  \frac{\partial^2 k(\mathbf{x}^*,{\mathbf{x}^*}')}{\partial x^*_d \partial {x^*_d}'} -
  \frac{\partial K(\mathbf{x}^*,\mathbf{X})}{\partial x^*_d} \left(K(\mathbf{X},\mathbf{X}) + \sigma^2 \mathbf{I}\right)^{-1}\frac{\partial K(\mathbf{X}, \mathbf{x}^*)}{\partial x^*_d}.
\end{equation*}

In the present application, the vector $\mathbf{x}$ includes coordinates of the atoms and the function $f$ is the energy of the system.
The extended observation vector
\begin{equation*}
\mathbf{y}_{\mathrm{ext}} = \begin{bmatrix}
            y^{(1)} \cdots y^{(N)},
            \frac{\partial f(\mathbf{x}^{(1)})}{\partial x^{(1)}_1}  \cdots \frac{\partial f(\mathbf{x}^{(N)})}{\partial x^{(N)}_1},
            \frac{\partial f(\mathbf{x}^{(1)})}{\partial x^{(1)}_2}  \cdots \frac{\partial f(\mathbf{x}^{(N)})}{\partial x^{(N)}_2}, & \ldots&,
            \frac{\partial f(\mathbf{x}^{(1)})}{\partial x^{(1)}_D}  \cdots \frac{\partial f(\mathbf{x}^{(N)})}{\partial x^{(N)}_D}
            \end{bmatrix}^{\mathsf{T}}
\end{equation*}
includes the accurate values of the energy and the partial derivatives of the energy with respect to the coordinates of the atoms (i.e., components of the negative force vector)
at the training data points $\mathbf{x}^{(1)}, \mathbf{x}^{(2)}, \ldots , \mathbf{x}^{(N)}$.
The GP model is used to predict the energy $f(\mathbf{x}^*)$ and its gradient vector
$\left[ \frac{\partial f(\mathbf{x}^*)}{\partial x^*_1}, \frac{\partial f(\mathbf{x}^*)}{\partial x^*_2}, \ldots, \frac{\partial f(\mathbf{x}^*)}{\partial x^*_D} \right]^\mathsf{T}$ at a new point $\mathbf{x}^*$, which in this case represents an image
on the discrete path representation between the initial and final state minima.
Since the training data also include derivative observations, the mean and variance of the
posterior predictive distribution of the energy are given as
\begin{equation}\label{gpmean}
  {\mathrm{E}}[f(\mathbf{x}^*)|\mathbf{y}_{\mathrm{ext}},\mathbf{X},\bm{\theta}] = {\mathbf{K}}_{\mathrm{ext}}^* \left({\mathbf{K}}_{\mathrm{ext}} + \sigma^2 \mathbf{I}\right)^{-1} \mathbf{y}_{\mathrm{ext}}
\end{equation}
and
\begin{equation}\label{gpvar}
  {\mathrm{Var}}[f(\mathbf{x}^*)|\mathbf{y}_{\mathrm{ext}},\mathbf{X},\bm{\theta}] =
  k(\mathbf{x}^*,\mathbf{x}^*) - {\mathbf{K}}_{\mathrm{ext}}^* \left({\mathbf{K}}_{\mathrm{ext}} + \sigma^2 \mathbf{I}\right)^{-1} {{\mathbf{K}}_{\mathrm{ext}}^*}^\mathsf{T},
\end{equation}
where
\begin{equation*}
{\mathbf{K}}_{\mathrm{ext}}^* = \begin{bmatrix}
            K(\mathbf{x}^*,\mathbf{X}) & \frac{\partial K(\mathbf{x}^*,\mathbf{X})}{\partial x_1} & \frac{\partial K(\mathbf{x}^*,\mathbf{X})}{\partial x_2} & \cdots & \frac{\partial K(\mathbf{x}^*,\mathbf{X})}{\partial x_D}
            \end{bmatrix}
\end{equation*}
and
\begin{equation*}
{\mathbf{K}}_{\mathrm{ext}} = \begin{bmatrix}
            K(\mathbf{X},\mathbf{X}) & \frac{\partial K(\mathbf{X},\mathbf{X}')}{\partial x_1'} & \frac{\partial K(\mathbf{X},\mathbf{X}')}{\partial x_2'} & \cdots & \frac{\partial K(\mathbf{X},\mathbf{X}')}{\partial x_D'} \\[0.3em]
            \frac{\partial K(\mathbf{X},\mathbf{X}')}{\partial x_1} & \frac{\partial^2 K(\mathbf{X},\mathbf{X}')}{\partial x_1 \partial x_1'}    & \frac{\partial^2 K(\mathbf{X},\mathbf{X}')}{\partial x_1 \partial x_2'}  & \cdots & \frac{\partial^2 K(\mathbf{X},\mathbf{X}')}{\partial x_1 \partial x_D'} \\[0.3em]
            \frac{\partial K(\mathbf{X},\mathbf{X}')}{\partial x_2} & \frac{\partial^2 K(\mathbf{X},\mathbf{X}')}{\partial x_2 \partial x_1'}    & \frac{\partial^2 K(\mathbf{X},\mathbf{X}')}{\partial x_2 \partial x_2'}  & \cdots & \frac{\partial^2 K(\mathbf{X},\mathbf{X}')}{\partial x_2 \partial x_D'} \\[0.3em]
            \vdots                                                  & \vdots                                                            & \vdots                                                          & \ddots & \vdots \\[0.3em]
            \frac{\partial K(\mathbf{X},\mathbf{X}')}{\partial x_D} & \frac{\partial^2 K(\mathbf{X},\mathbf{X}')}{\partial x_D \partial x_1'}    & \frac{\partial^2 K(\mathbf{X},\mathbf{X}')}{\partial x_D \partial x_2' } & \cdots & \frac{\partial^2 K(\mathbf{X},\mathbf{X}')}{\partial x_D \partial x_D'}
          \end{bmatrix}.
\end{equation*}
Correspondingly, the mean and variance of the posterior predictive distribution of the partial derivative of the energy
with respect to coordinate $d$ at $\mathbf{x^*}$ are given as
\begin{equation}\label{gradmean}
  {\mathrm{E}}\left[ \frac{\partial f(\mathbf{x^*})}{\partial x^*_d} \middle| \mathbf{y}_{\mathrm{ext}},\mathbf{X},\bm{\theta} \right] =
  \frac{\partial {\mathbf{K}}_{\mathrm{ext}}^*}{\partial x^*_d} \left({\mathbf{K}}_{\mathrm{ext}} + \sigma^2 \mathbf{I}\right)^{-1} \mathbf{y}_{\mathrm{ext}}
\end{equation}
and
\begin{equation*}
  {\mathrm{Var}}\left[ \frac{\partial f(\mathbf{x^*})}{\partial x^*_d} \middle| \mathbf{y}_{\mathrm{ext}},\mathbf{X},\bm{\theta} \right] =
  \frac{\partial^2 k(\mathbf{x}^*,{\mathbf{x}^*}')}{\partial x_d^* \partial {x_d^*}'} -
  \frac{\partial {\mathbf{K}}_{\mathrm{ext}}^*}{\partial x^*_d} \left({\mathbf{K}}_{\mathrm{ext}} + \sigma^2 \mathbf{I}\right)^{-1} \frac{\partial {{\mathbf{K}}_{\mathrm{ext}}^*}^{\mathsf{T}}}{\partial x^*_d}.
\end{equation*}
Even if the observations are assumed to be accurate, a small but positive value for the noise variance $\sigma^2$
is used to avoid numerical problems when inverting the covariance matrix ${\mathbf{K}}_{\mathrm{ext}}$.


\section{Algorithms}
\label{sec:Algorithms}

In this section, the two algorithms using the Gaussian process regression (GPR) approach to accelerate CI-NEB calculations,
the all-images-evaluated (AIE) algorithm and the one-image-evaluated (OIE) algorithm, are presented in detail.

\vskip 0.3 true cm

\begin{list}{}{\topsep=0in \leftmargin=0.2in \rightmargin=0.2in}

\item {\bf All-images-evaluated (AIE) algorithm}
\vskip 0.2 true cm

\item {\bf Input:} a GP model, energy (and zero force) at the two minima on the energy surface,
coordinates of the $N_\mathrm{im}$ images on the initial path,
a final convergence threshold $T_\mathrm{MEP}$ for the minimum energy path,
an additional final convergence threshold $T_\mathrm{CI}$ for the climbing image $i_\mathrm{CI}$,
a preliminary convergence threshold $T_\mathrm{CIon}^\mathrm{GP}$
for turning climbing image mode on during the relaxation phase,
a maximum displacement $r_\mathrm{max}$ of any image from the nearest observed data point.

\vskip 0.2 true cm

\item {\bf Output:} a minimum energy path represented by $N_\mathrm{im}$ images
one of which has climbed to the highest saddle point.
\vskip 0.2 true cm

\item 1. Start from the initial path, and repeat the following (outer iteration loop):
\vskip 0.2 true cm

\end{list}

  \begin{enumerate}[leftmargin=0.7in, rightmargin=0.2in, label={\Alph*.}]
  \item Evaluate the true energy and force at the $N_\mathrm{im}-2$ intermediate images of the current path, and add them to the training data.
  \item Calculate the accurate NEB force vector $F_\mathrm{NEB}(i)$ for each intermediate image $i \in \{2,3,\ldots,$ $N_\mathrm{im}-1$\}.
  \item If $\max_i |F_\mathrm{NEB}(i)| < T_\mathrm{MEP}$ and $|F_\mathrm{NEB}(i_\mathrm{CI})| < T_\mathrm{CI}$, then stop the algorithm (final convergence reached).
  \item Optimize the hyperparameters of the GP model based on the training data, and calculate the matrix inversion in Eq.~\ref{gpmean}.
  \item Relaxation phase: Start from the initial path, set climbing image mode off, and repeat the following (inner iteration loop):
    \begin{enumerate}[label={\Roman*.}]
    \item Calculate the GP posterior mean energy and gradient at the intermediate images using Eqs.~\ref{gpmean} and \ref{gradmean}.
    \item Calculate the approximate NEB force vector $F_\mathrm{NEB}^\mathrm{GP}(i)$ for each intermediate image using the GP posterior mean gradient.
    \item If climbing image mode is off and  $\max_i |F_\mathrm{NEB}^\mathrm{GP}(i)| < T_\mathrm{CIon}^\mathrm{GP}$, then turn climbing image mode on and recalculate the approximate NEB force vectors.
    \item If climbing image mode is on and $\max_i |F_\mathrm{NEB}^\mathrm{GP}(i)| < \frac{1}{10} T_\mathrm{CI}$, then stop the relaxation phase (E).
    \item Move the intermediate images along the approximate NEB force vector $F_\mathrm{NEB}^\mathrm{GP}(i)$ with a step size defined by the projected velocity Verlet algorithm.
    \item If the distance from any current image to the nearest observed data point is larger than $r_\mathrm{max}$, then reject the last inner step and stop the relaxation phase (E).
    \end{enumerate}
  \end{enumerate}
  
\vskip 0.2 true cm

\begin{figure}[htbp]
\centering
\includegraphics[width=\columnwidth]{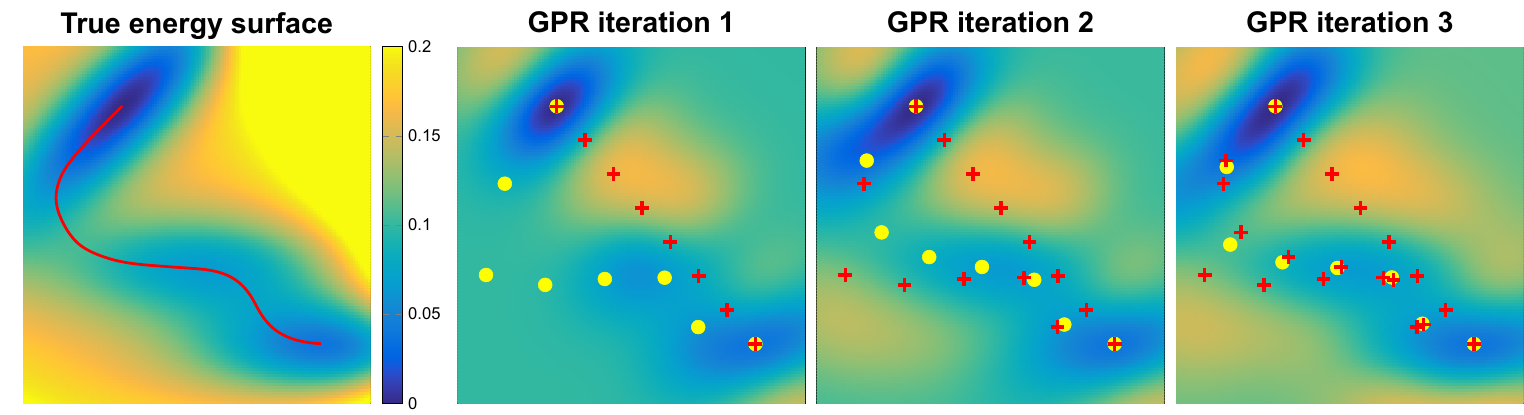}
\caption{
Far left: The two-dimensional M\"uller-Brown energy surface, which has three minima, and the minimum energy path (red curve).
Three panels to the right: An illustration of the iterative construction of an approximate energy surface 
in the vicinity of the minimum energy path using the all-images-evaluated algorithm where
energy and atomic forces are evaluated at all intermediate images of the nudged elastic band.
The initial path is a straight line interpolation between the initial and final state minima.
The red + signs show the points at which the energy
and atomic forces have been evaluated. 
The yellow disks show the climbing image nudged elastic band relaxed on the approximate energy surface of each Gaussian process regression iteration.
After each GPR iteration, final convergence of the path is checked by energy and force evaluations, which are then added
to the training data for the following GPR iteration.
After three iterations (and a total of 24 energy and force evaluations),
final convergence is confirmed as the magnitude of the NEB force is below the threshold 0.01 both for
the climbing image and the other intermediate images.
}
\label{fig:fig1}
\end{figure}

A pseudocode for the AIE algorithm is presented above.
Figure~\ref{fig:fig1} shows an illustration of the progression of the algorithm on the two-dimensional M\"uller-Brown energy surface \cite{MullerBrown79}.
The energy (and the zero gradient of the energy) at the initial and final state minima are assumed to be provided as input.

\begin{figure}[htbp]
\centering
\includegraphics[width=\columnwidth]{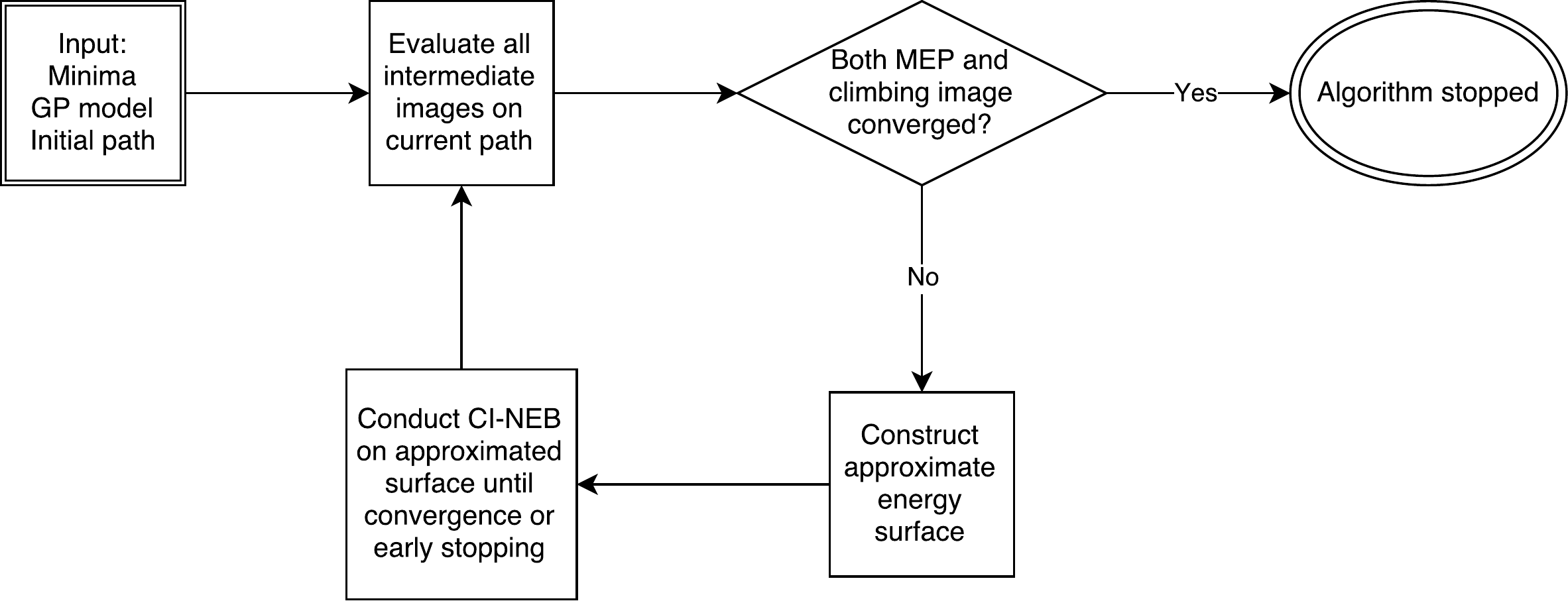}
\caption{
A flowchart of the all-images-evaluated algorithm, where energy and atomic forces are evaluated at all intermediate images of the climbing image nudged elastic band
relaxed on the GP-approximated energy surface.
}
\label{fig:fig2}
\end{figure}

The algorithm is started by evaluating the energy and force at the $N_\mathrm{im} - 2$ intermediate images of an initial path
and constructing a GP model for the energy based on the obtained information.
The path is then relaxed on the approximate energy surface (GPR iteration 1), which is given as the posterior mean of the GP model,
with a regular CI-NEB method using the posterior mean gradient of the GP model
to calculate the approximate NEB force at the images.
After each relaxation phase, final convergence of the path is checked by evaluating the true energy and force
at the images of the relaxed path, and these observations are then added to the training data to improve the
GP model on the following round.
As can be seen from Fig.~\ref{fig:fig1}, a fairly accurate approximation of the M\"uller-Brown energy surface is obtained already after three GPR
iterations of the AIE algorithm. This corresponds to 18 energy and force evaluations since the path is represented by six
movable images in this case.
A simplified flowchart of the algorithm is presented in Fig.~\ref{fig:fig2}.

The final convergence of the climbing image nudged elastic band is defined by
the magnitude of the NEB force (including the spring force parallel to the path tangent)
at each image calculated using the energy and force of the true energy surface.
Two separate final convergence thresholds are used: $T_\mathrm{MEP}$ for the maximum
NEB force magnitude $\max_i |F_\mathrm{NEB}(i)|$ among the intermediate images $i$ and
a tighter $T_\mathrm{CI}$ for the NEB force magnitude $|F_\mathrm{NEB}(i_\mathrm{CI})|$
of the climbing image $i_\mathrm{CI}$.

To ensure that the incomplete relaxation of the path on the approximate energy surface
does not disturb final convergence,
the convergence threshold $T_\mathrm{MEP}^\mathrm{GP}$ for the maximum approximate NEB force magnitude
$\max_i |F_\mathrm{NEB}^\mathrm{GP}(i)|$ during the relaxation phase is
defined as 1/10 of the tighter final threshold $T_\mathrm{CI}$.
To decrease the amount of inner iterations during the relaxation phase,
there is an alternative option for $T_\mathrm{MEP}^\mathrm{GP}$ to be defined
as 1/10 of the smallest true NEB force magnitude obtained so far on any of the intermediate images
(but not less than $T_\mathrm{CI}/10$).
If the approximation error is assumed not to decrease more than that during one GPR iteration,
there is no need for a tighter convergence on the approximated surface and thus the relaxation phase
can be sped up by using a larger tolerance.
The divisor 10 can also be replaced by some other suitable number.

To make it more certain that the path converges to the same MEP as the one obtained by the regular CI-NEB method,
each relaxation phase on the approximate energy surface is started from the same initial path.
The relaxation is first conducted without climbing image mode until a preliminary convergence threshold
$T_\mathrm{CIon}^\mathrm{GP}$ is reached and then continued from the preliminary evenly spaced path
with climbing image mode turned on.
Starting each relaxation phase from the initial path may possibly improve the stability of the algorithm,
but there is also an alternative option which may decrease the number of inner iterations.
In this alternative scheme, each relaxation phase would be started from the latest evenly spaced path converged to $T_\mathrm{CIon}^\mathrm{GP}$,
and the climbing image phase would be started from the latest converged CI-NEB path if the climbing image of that path
has the highest energy also on the current approximate energy surface. If, instead, the index of the highest energy image
has changed, the climbing image phase would be started normally from the preliminarily relaxed evenly spaced path.

In the early phases of the algorithm, when little information is available about the energy surface,
there is a greater possibility that the path wanders far away from the initial path.
To prevent this behavior, it is good to have some early stopping rule for the movement of the path.
Thus, if the distance from any current image to the nearest observed data point becomes larger than $r_\mathrm{max}$,
then the last inner step is rejected and the relaxation phase stopped.
In the heptamer island benchmark described later in Sec.~\ref{sec:Application}, $r_\mathrm{max}$ is
defined as half of the length of the initial path (sum of the distances between adjacent images),
but other definitions, e.g., based on the length scale of the GP model, are also possible.

When the final goal is to estimate
the transition rates using harmonic transition state theory,
the Hessian matrices at the initial and final state minima will need to be calculated.
The Hessian is usually calculated with a finite difference method,
where energy and force evaluations are made in the neighborhood of the minima.
If these calculations, which anyway are needed for the Hessian, are evaluated already in the beginning of the MEP calculation,
the calculated values can be added to the initial data set for the GPR calculations
to provide information about the shape of the energy surface around the endpoints of the path
and improve especially the early phase of the algorithm.
To test the effect of the Hessian input in the heptamer island benchmark,
a finite difference displacement of $10^{-3}$~{\AA} is made in
the positive direction along each of the atom coordinates, and the values of the energy and force
at these points are included as input in the GPR calculation.

Since the GPR calculations using gradient observations require an inversion of an $(1+D)N \times (1+D)N$ matrix,
the computational effort scales as $\mathcal{O}(((1+D)N)^3)$,
where $N$ is the number of observation points and $D$ is the number of degrees of freedom (here coordinates of movable atoms).
As usual, the matrix inversion is computed by forming a Cholesky decomposition and solving a linear system of equations.
Since the model stays the same during the relaxation phase on the approximate energy surface,
the matrix inversion needs to be computed only once for each GPR iteration and
the complexity of one inner iteration on the approximate energy surface is $\mathcal{O}((1+D)N)$.

\vskip 0.3 true cm

\begin{list}{}{\topsep=0in \leftmargin=0.2in \rightmargin=0.2in}

\item {\bf One-image-evaluated (OIE) algorithm}
\vskip 0.2 true cm

\item {\bf Input:} a GP model, energy (and zero force) at the two minima on the energy surface,
coordinates of the $N_\mathrm{im}$ images on the initial path,
a final convergence threshold $T_\mathrm{MEP}$ for the minimum energy path,
an additional final convergence threshold $T_\mathrm{CI}$ for the climbing image $i_\mathrm{CI}$,
a preliminary convergence threshold $T_\mathrm{CIon}^\mathrm{GP}$
for turning climbing image mode on during the relaxation phase,
a maximum displacement $r_\mathrm{max}$ of any image from the nearest observed data point.

\vskip 0.2 true cm

\item {\bf Output:} a minimum energy path represented by $N_\mathrm{im}$ images
one of which has climbed to the highest saddle point.
\vskip 0.2 true cm

\item 1. Optimize the hyperparameters of the GP model based on the initial data.
\vskip 0.2 true cm

\item 2. Start from the initial path, and repeat the following (outer iteration loop):
\vskip 0.2 true cm

\end{list}

  \begin{enumerate}[leftmargin=0.7in, rightmargin=0.2in, label={\Alph*.}]
  \item Calculate the GP posterior variance at the unevaluated images $i \in \mathrm{I}_\mathrm{u}$ on the current path using Eq.~\ref{gpvar}.
  \item Evaluate the true energy and force at the image with highest posterior variance, and add them to the training data.
  \item Calculate the accurate NEB force vector $F_\mathrm{NEB}(i)$ for the evaluated images $i \in \mathrm{I}_\mathrm{e}$.
  \item If all images on the current path have been evaluated, $\max_i |F_\mathrm{NEB}(i)| < T_\mathrm{MEP}$ and $|F_\mathrm{NEB}(i_\mathrm{CI})| < T_\mathrm{CI}$, then stop the algorithm (final convergence reached).
  \item Reoptimize the GP hyperparameters, calculate the GP posterior mean energy and gradient at the unevaluated images $i \in \mathrm{I}_\mathrm{u}$ using Eqs.~\ref{gpmean} and \ref{gradmean}, and save the matrix inversion for further use.
  \item Calculate the approximate NEB force vector $F_\mathrm{NEB}^\mathrm{GP}(i)$ for the unevaluated images $i \in \mathrm{I}_\mathrm{u}$ using the GP posterior mean gradient, and set $F_\mathrm{NEB}^\mathrm{GP}(i) = F_\mathrm{NEB}(i)$ for the evaluated images $i \in \mathrm{I}_\mathrm{e}$.
  \item If $\max_i |F_\mathrm{NEB}^\mathrm{GP}(i)| < T_\mathrm{MEP}$:
    \begin{enumerate}[label={\Roman*.}]
    \item If $i_\mathrm{CI} \in I_\mathrm{u}$, then evaluate the energy and force at the climbing image $i_\mathrm{CI}$, add them to the training data and go to C.
    \item If $i_\mathrm{CI} \in I_\mathrm{e}$ and $|F_\mathrm{NEB}(i_\mathrm{CI})| < T_\mathrm{CI}$, then go to A.
    \item If $i_\mathrm{CI} \in I_\mathrm{e}$ and $|F_\mathrm{NEB}(i_\mathrm{CI})| \geq T_\mathrm{CI}$, then execute the relaxation phase (H), evaluate the energy and force at the climbing image, add them to the training data and go to C.
    \end{enumerate}
  \item Relaxation phase: Start from the initial path, set climbing image mode off, and repeat the following (inner iteration loop):
    \begin{enumerate}[label={\Roman*.}]
    \item Calculate the GP posterior mean energy and gradient at the intermediate images using Eqs.~\ref{gpmean} and \ref{gradmean}.
    \item Calculate the approximate NEB force vector $F_\mathrm{NEB}^\mathrm{GP}(i)$ for each intermediate image using the GP posterior mean gradient.
    \item If climbing image mode is off and $\max_i |F_\mathrm{NEB}^\mathrm{GP}(i)| < T_\mathrm{CIon}^\mathrm{GP}$, then turn climbing image mode on and recalculate the approximate NEB force vectors.
    \item If climbing image mode is on and $\max_i |F_\mathrm{NEB}^\mathrm{GP}(i)| < T_\mathrm{MEP}^\mathrm{GP} = \frac{1}{10} T_\mathrm{CI}$, then stop the relaxation phase (H).
    \item Move the intermediate images along the approximate NEB force vector $F_\mathrm{NEB}^\mathrm{GP}(i)$ with a step size defined by the projected velocity Verlet algorithm.
    \item If the distance from any current image $i$ to the nearest observed data point is larger than $r_\mathrm{max}$, then reject the last inner step, evaluate image $i$ and go to C.
    \end{enumerate}
  \end{enumerate}
  
\vskip 0.2 true cm

\begin{figure}[htbp]
\centering
\includegraphics[width=\columnwidth]{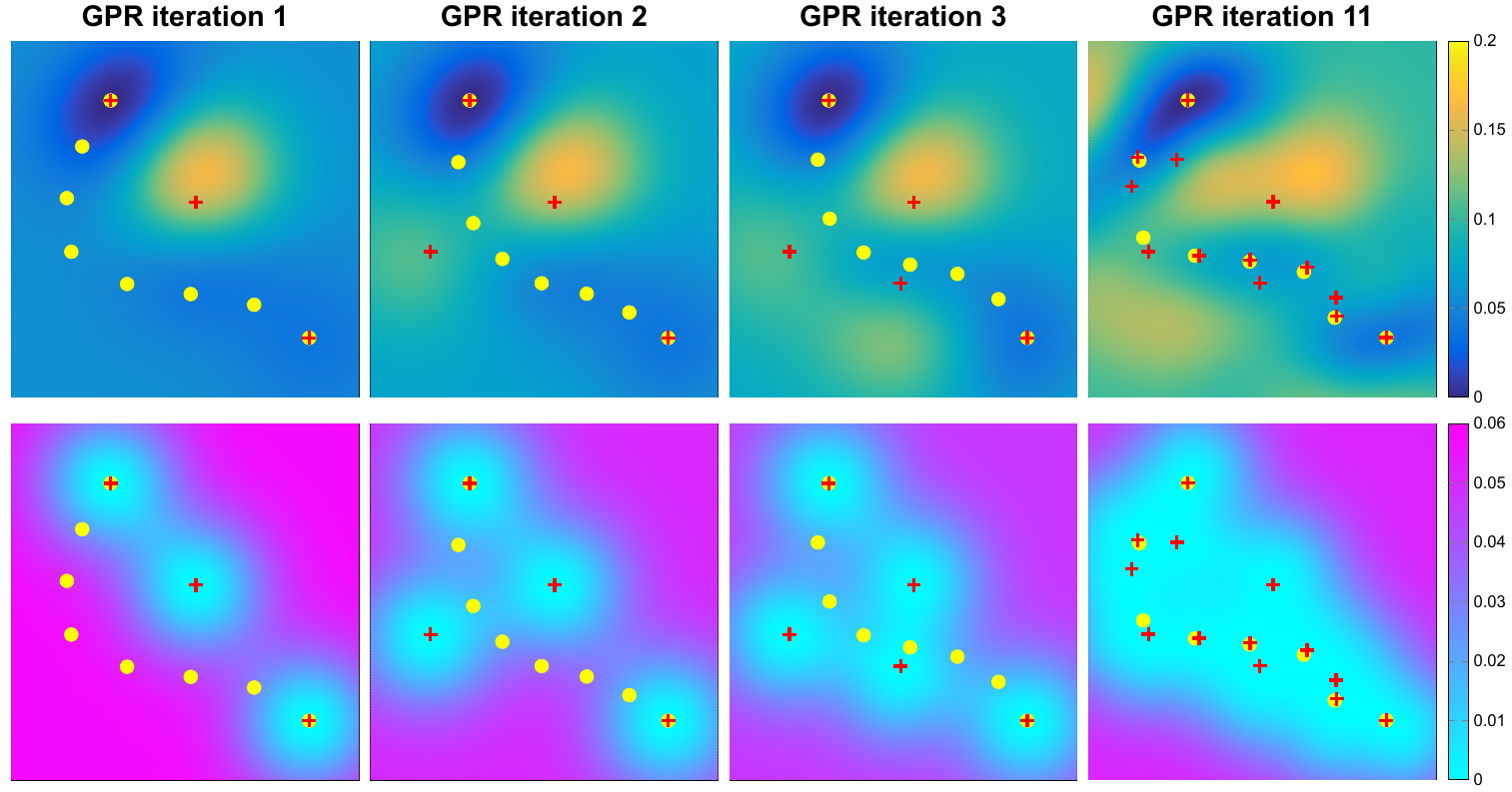}
\caption{
An illustration of the iterative construction of an approximate energy surface to the
two-dimensional M\"uller-Brown energy surface (shown in Fig.~\ref{fig:fig1})
in the vicinity of the minimum energy path
using the one-image-evaluated algorithm where the energy and atomic forces are evaluated
only at one image of the nudged elastic band.
The initial path is a straight line interpolation between the initial and final state minima.
Upper panel:
The red + signs show the points at which the energy and atomic forces have been evaluated.
The yellow disks show the climbing image nudged elastic band
relaxed on the approximate energy surface of each Gaussian process regression iteration.
After GPR iterations 1, 2 and 3, the energy and force are calculated at the image where the estimated uncertainty
is largest and the observed data are then added to the training data set for the following GPR iteration.
Lower panel:
The estimated uncertainty (standard deviation) of the energy approximation shown directly above.
After eleven iterations, the path is not displaced further but the final convergence
is checked by evaluating the energy and force at each intermediate image one by one.
After 17 evaluations, final convergence is confirmed as the magnitude of the NEB force is
below the threshold 0.01 both for the climbing image and the other intermediate images.
}
\label{fig:fig3}
\end{figure}

A pseudocode for the OIE algorithm is presented above.
Figure~\ref{fig:fig3} shows an illustration of the progression of the algorithm on the M\"uller-Brown energy surface.
Both the GP approximation to the energy surface and the estimated uncertainty after
one, two, three and eleven GPR iterations are shown.
The energy (and the zero gradient of the energy) at the initial and final state minima are assumed to be given as input.

The algorithm is started by evaluating the true energy and force at the image located in the middle of the initial path
where the uncertainty of the initial GP model is largest. The GP model is then updated based on the obtained information,
and the path is relaxed on the approximate energy surface (GPR iteration 1).
After each relaxation phase, the true energy and force are evaluated at only one image of the relaxed path before updating the GP model.
According to the main rule, the image with the highest uncertainty estimate is selected for evaluation,
and the information obtained is then used to improve the GP model on the following round.
As can be seen from Fig.~\ref{fig:fig3}, a fairly accurate approximation of the M\"uller-Brown energy surface
is obtained already after eleven GPR iterations of the OIE algorithm.
This corresponds to only eleven energy and force evaluations, quite a bit fewer than the 18 included in the three GPR iterations of the
AIE algorithm shown in Fig.~\ref{fig:fig1}.

The details of the OIE algorithm are otherwise similar to the AIE algorithm, but since only one image is evaluated between the GPR iterations,
relaxing the path between each evaluation would mean that the accurate NEB force is only known for one image at a time.
Thus, it would not be known for sure whether the path has converged on a true MEP.
Approximations for the NEB forces can of course be calculated at the unevaluated images based on the updated GP model, but
since the NEB forces have been relaxed to zero based on the previous GP approximation and since the largest changes to
the approximation usually emerge near the new observation point, it is most likely that these approximations underestimate the NEB force magnitudes.
The approximated NEB forces, however, together with the accurate ones, at least indicate if there is a possibility
that the path may have converged.
The general idea for the convergence check of the OIE algorithm is that when the maximum magnitude of both the accurate and
approximated NEB forces is below the final convergence threshold $T_\mathrm{MEP}$, more images are evaluated without moving the path
until some of the magnitudes rise above the threshold or all images have been evaluated.
Since the images with the highest uncertainty, which are the most likely ones to violate the convergence criterion, are evaluated first,
it is likely that the convergence check will be interrupted early if the path has not yet truly converged.

The special role of the climbing image makes the evaluation rules a bit more complicated.
Since the climbing image has a tighter convergence threshold
and since the position of the climbing image affects also the distribution of the other images,
it is desirable to favor evaluations of the climbing image during the convergence check.
As long as the maximum magnitude of the accurate and approximated NEB forces is above $T_\mathrm{MEP}$,
the GP relaxation phase is executed normally and the image with the highest uncertainty is evaluated.
After the maximum magnitude has reduced below $T_\mathrm{MEP}$, the climbing image is evaluated without moving the path (if not already evaluated).
As long as the maximum NEB force magnitude stays below $T_\mathrm{MEP}$
but the NEB force magnitude $|F_\mathrm{NEB}(i_\mathrm{CI})|$ on the
climbing image is above $T_\mathrm{CI}$, the path is relaxed and the climbing image re-evaluated.
Finally, if the maximum magnitude of the accurate and approximated NEB forces is below $T_\mathrm{MEP}$,
the climbing image has been evaluated and $|F_\mathrm{NEB}(i_\mathrm{CI})| < T_\mathrm{CI}$,
then more images are evaluated without moving the path, starting from the image with the highest uncertainty.

\begin{figure}[htbp]
\centering
\includegraphics[width=\columnwidth]{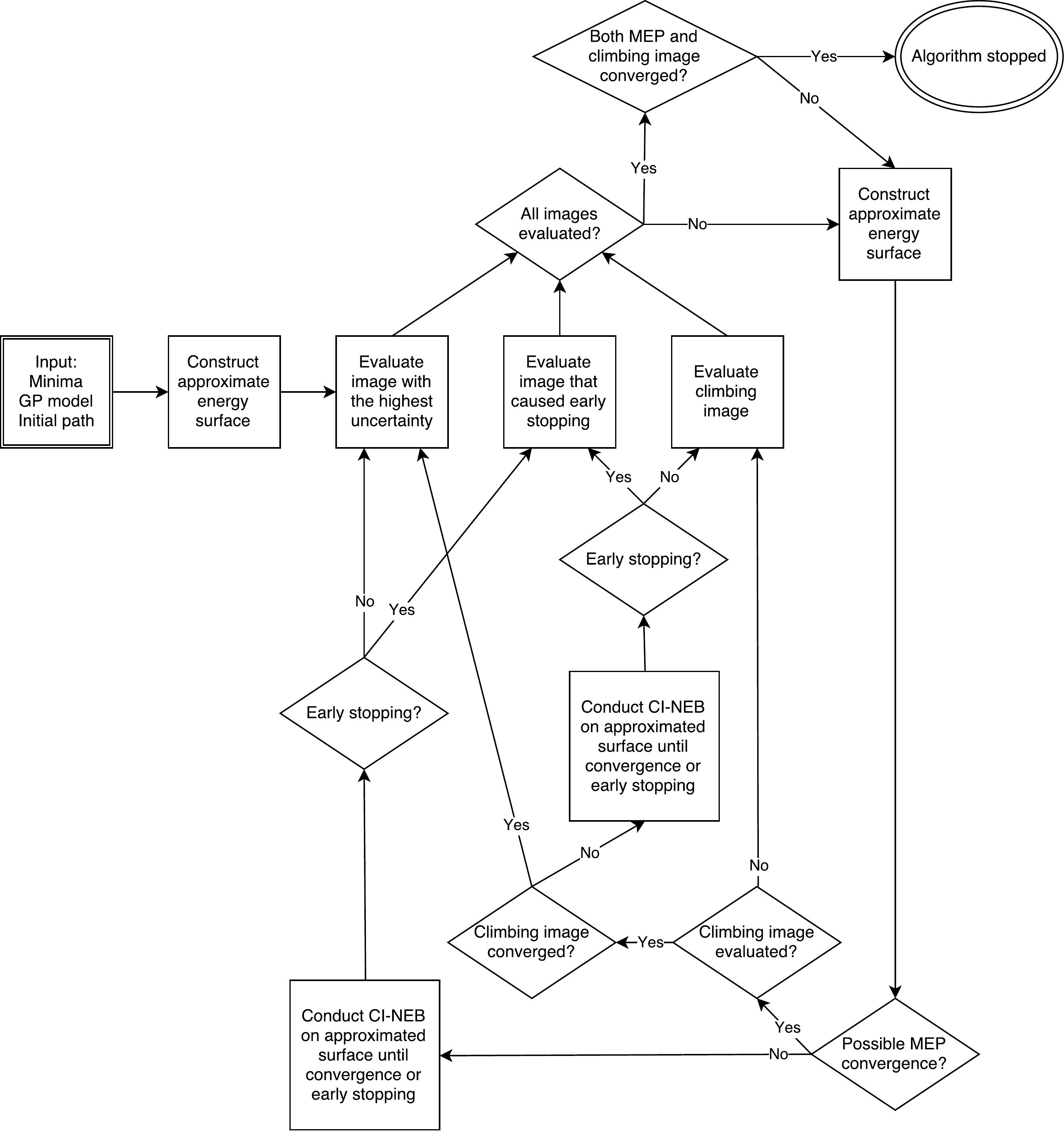}
\caption{
A flowchart of the one-image-evaluated algorithm, where the energy and atomic forces are evaluated at only one image before a new Gaussian process regression iteration.
}
\label{fig:fig4}
\end{figure}

Another exception to the selection of the image to be evaluated is caused by the early stopping rule during the relaxation phase.
If the distance from any current image $i$ to the nearest observed data point becomes larger than $r_\mathrm{max}$,
then the last inner step is rejected, the relaxation phase stopped and image $i$ evaluated next.
A simplified flowchart of the OIE algorithm is presented in Fig.~\ref{fig:fig4}.

\section{Application to the heptamer island benchmark}
\label{sec:Application}

\begin{figure}[htbp]
\centering
\includegraphics[width=.5 \columnwidth]{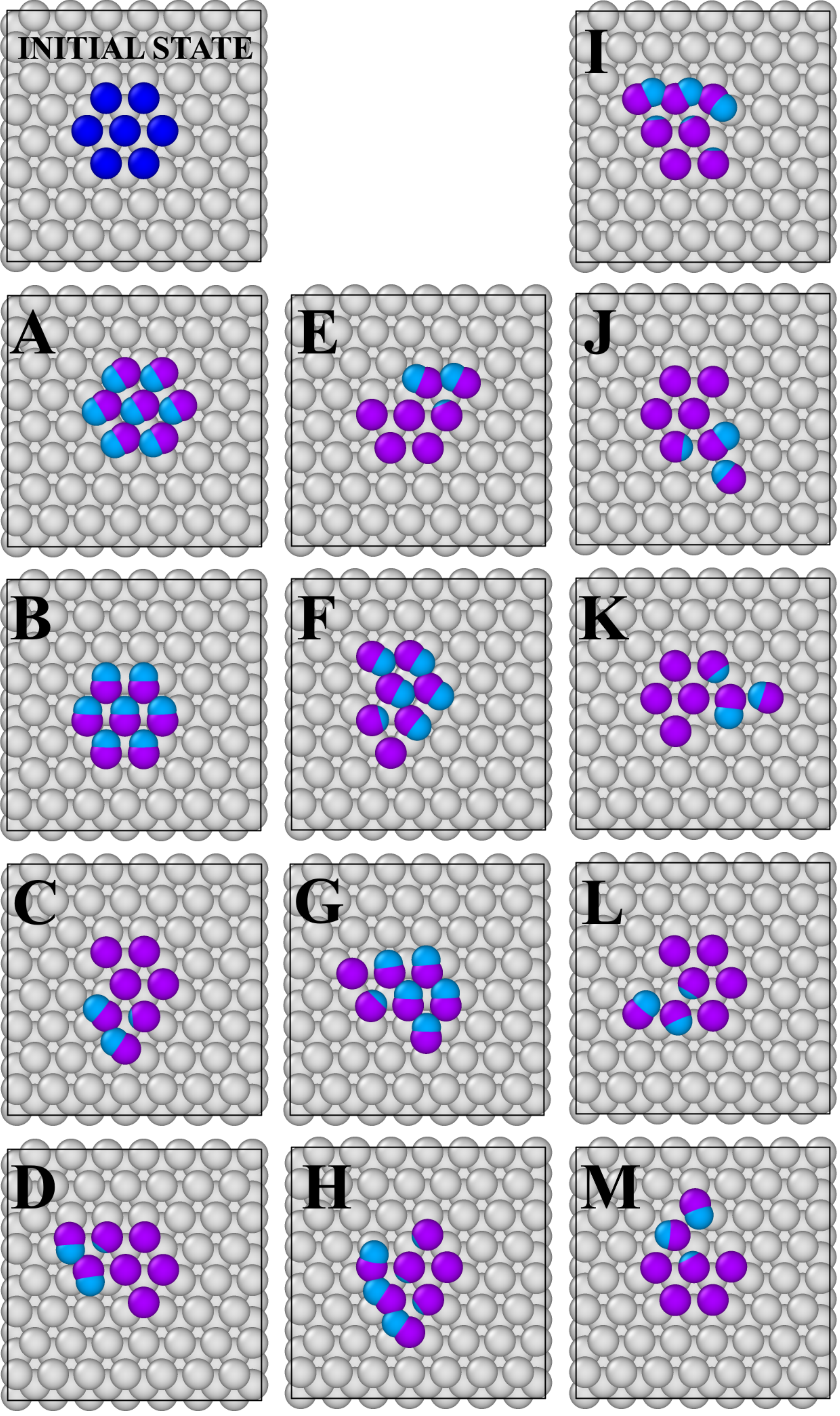}
\caption{
An illustration of the atomic transitions of the heptamer island benchmark problem.
The initial configuration involves a seven-atom island (dark blue disks in uppermost left column) 
adsorbed at FCC sites of an FCC(111) surface.
Saddle point configuration (light blue disks) and final configuration (purple disks) are shown together
for each of the transitions, labeled $A$ -- $M$.
}
\label{fig:fig5}
\end{figure}

A test problem that has been used in several studies of algorithms for finding MEPs and saddle points involves an
island of seven atoms on the (111) surface of a face-centered cubic (FCC) crystal \cite{Henkelman00,Chill2014}.
The interaction between the atoms is described with a simple Morse potential to make the implementation of the benchmark easy.
The initial, saddle point and final configurations of the atoms for the 13 lowest activation energy transitions, 
labeled from $A$ to $M$, are shown in Fig.~\ref{fig:fig5}.
In the initial state, the seven atoms sit at FCC surface sites and form a compact island.
In transitions $A$ and $B$, the whole island shifts to HCP sites on the surface.  
In some of the other transitions, a pair of edge atoms slides to adjacent FCC sites,
an atom half way dissociates from the island, or a pair of edge atoms moves in such a way that one of the atoms is displaced away from the island while the other atom takes its place.

The calculations were carried out using five intermediate images ($N_\mathrm{im}=7$) in the CI-NEB calculations starting with an IDPP path,
and the images were moved iteratively to an MEP using the projected velocity Verlet algorithm \cite{NEBleri} with a time step of 0.1~fs.
A time step of 1~fs is too large and leads to overshooting, but a time step of 0.01~fs requires a significantly larger number of iterations.  
The algorithms were continued until the magnitude of the true NEB force acting on the climbing image had dropped below $T_\mathrm{CI} = 0.01$~eV/{\AA}.
A larger tolerance, $T_\mathrm{MEP} = 0.3$ eV/{\AA}, was used for the magnitude of the NEB force acting on the other images in the CI-NEB calculation.
During each relaxation phase on the GP-approximated surface, the preliminary convergence threshold
$T_\mathrm{CIon}^\mathrm{GP}$ for turning climbing image mode on was 1 eV/{\AA}.
The GPR calculations were carried out using the GPstuff toolbox \cite{Vanhatalo13}.
The fixed parameters of the GP model were chosen to be $\sigma^2 = 10^{-8}$~eV$^2$ and $\sigma_\mathrm{c}^2 = 100$~eV$^2$.
A common length scale $l_d = l$ was used for all dimensions $d = 1, \ldots, D$,
and a zero mean Student's $t$-distribution (restricted to positive values) with 1~{\AA}$^2$ scale and four degrees of freedom
was used as a prior distribution for $l$ and a log-uniform distribution for $\sigma_{\mathrm{m}}^2$ (i.e., the default priors of GPstuff) in the optimization of the hyperparameters.
The prior distributions stabilize the point estimates of the hyperparameters especially in the beginning, when there is little data available.
The optimization was performed using the scaled conjugate gradient algorithm \cite{Bishop95}.

By using input from the Hessian at the initial and final state minima, the path relaxed on the GP-approximated energy surface can become qualitatively
similar to the true MEP with fewer GPR iterations.
This is illustrated for transition $I$ in Fig. 6. It shows
the true energy evaluated at the location of the images of the initial path, the converged MEP and the path after one GPR iteration
in the AIE algorithm with and without input from the Hessian.
The estimate without the Hessian input is quite poor at this point, the path reaching an area of high energy and the climbing
image moving to the final state minimum.

\begin{figure}[htbp]
\centering
\includegraphics{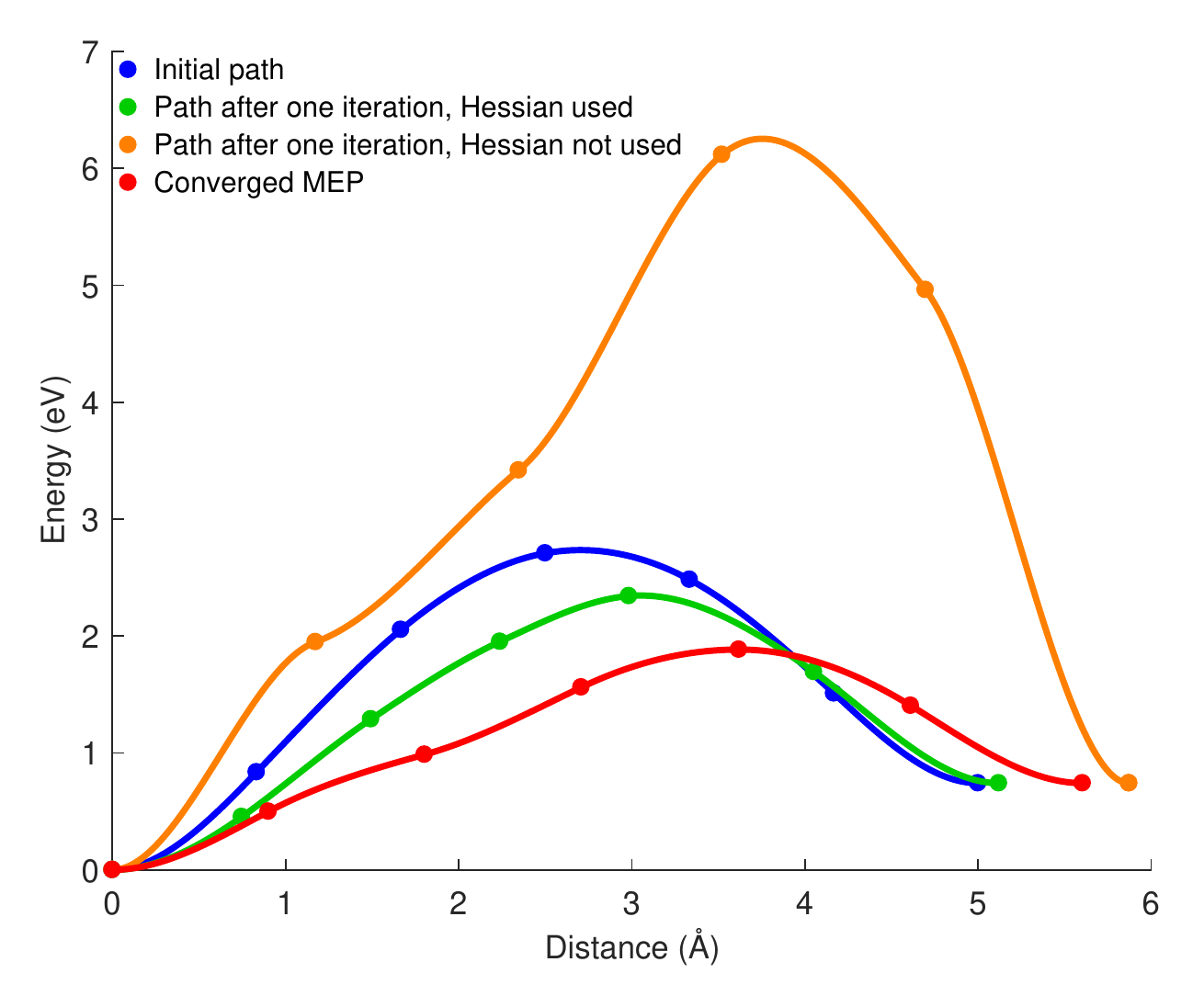}
\caption{
An illustration of the improvement that can be obtained by using the Hessian at the initial and final state minima.
The calculations are for transition $I$ when the substrate atoms are frozen.
The green and orange points show the true energy evaluated at the location of images of a climbing image nudged elastic band
relaxed on the first approximate energy surface of the all-images-evaluated
algorithm, i.e., at this point the energy and force had been evaluated at all the images of the initial path. 
The path obtained when the Hessian is not used (orange) is far from the 
converged minimum energy path (red), quite a bit farther away than the initial path (blue),
and the climbing image has moved to the final state minimum.
The path obtained when the Hessian is used (green), however, represents a clear improvement to the initial path,
having moved significantly closer to the converged MEP. 
Eventually, the path converges after 14 Gaussian process regression iterations
(a total of 75 energy and force evaluations required to confirm final convergence) when the Hessian is used,
and after 17 GPR iterations when it is not used.
}
\label{fig:fig6}
\end{figure}

\begin{figure}[htbp]
\centering
\includegraphics{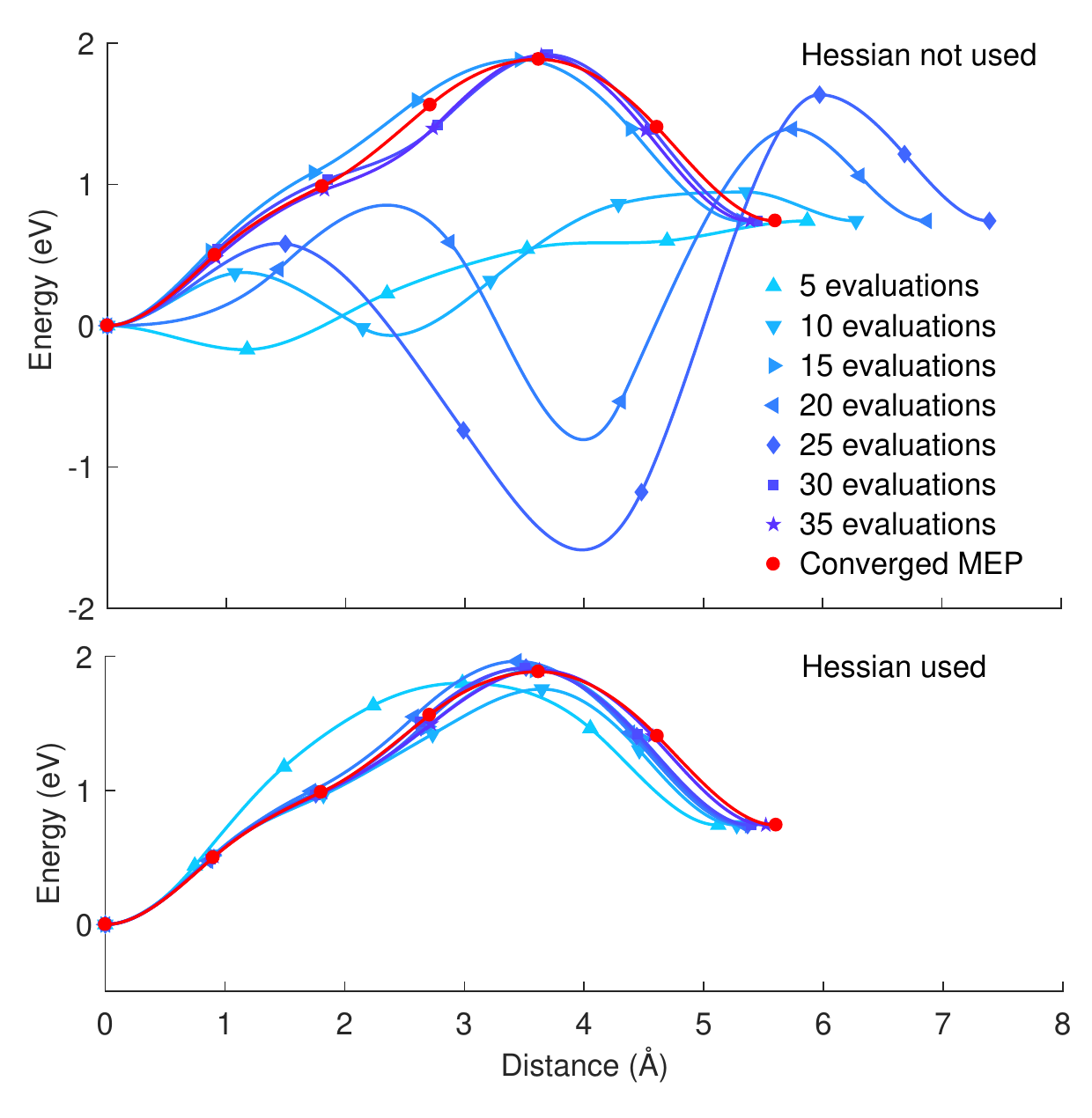}
\caption{
Comparison of the performance of the all-images-evaluated algorithm with and without input from the
Hessian at the initial state minima.
The markers show the GP-approximated energy at the images of the climbing image nudged elastic band
relaxed on the approximate energy surface after one to seven Gaussian process regression iterations (5 to 35 energy and force evaluations, blue)
and after final convergence (red) for transition $I$ when the substrate atoms are frozen.
Without the Hessian input (upper graph), the approximate energy surface after one GPR
iteration does not have an intermediate barrier and the climbing image moves to the final state minimum.
The true energy evaluated at each image along this path is shown in Fig.~\ref{fig:fig6} and shows a large energy barrier.
It takes six GPR iterations (30 energy and force evaluations) before the path relaxed on the approximated surface starts to
look qualitatively similar to the converged minimum energy path.
With the Hessian input (lower graph), the energy along the path relaxed on the approximate energy surface
is qualitatively similar to that of the converged MEP already after one iteration.
}
\label{fig:fig7}
\end{figure}

The reason for this behavior can be seen from Fig.~\ref{fig:fig7},
where the GP-approximated energy at the location of the images after one GPR iteration (five energy and force evaluations) is shown.
The maximum of the approximated energy along the path is indeed at the final state of the path.
However, with the input from the Hessian, a qualitatively correct path is obtained already after one GPR iteration. 
In this case the information coming from the Hessian about the curvature at the endpoints ensures that the GP-approximated
surface has minima at those points.
Without the Hessian input, the three subsequent GPR iterations still show qualitatively wrong variation of the energy along the
path, and it is only after the fifth iteration that enough input has been obtained for the GP model to be reasonably accurate
to produce a path qualitatively similar to the true MEP.

The overall reduction in the number of energy and force evaluations corresponds essentially to
the first three GPR iterations that can be skipped when the Hessian is provided. 
It takes 17 GPR iterations without and 14 iterations with the Hessian input to reach convergence in this case. 
The energy and force evaluations needed to construct the Hessian are not counted in the cost of finding the MEP since
they need to be carried out anyway if the transition rate is estimated using harmonic transition state theory.
While the reduction in the number of energy and force evaluations (here 15) corresponds to savings of about 20\%,
the importance of the Hessian input can be greater in more challenging systems when avoiding exploration of regions
far away from the MEP where the atomic forces tend to be large.

\begin{figure}[htbp]
\centering
\includegraphics{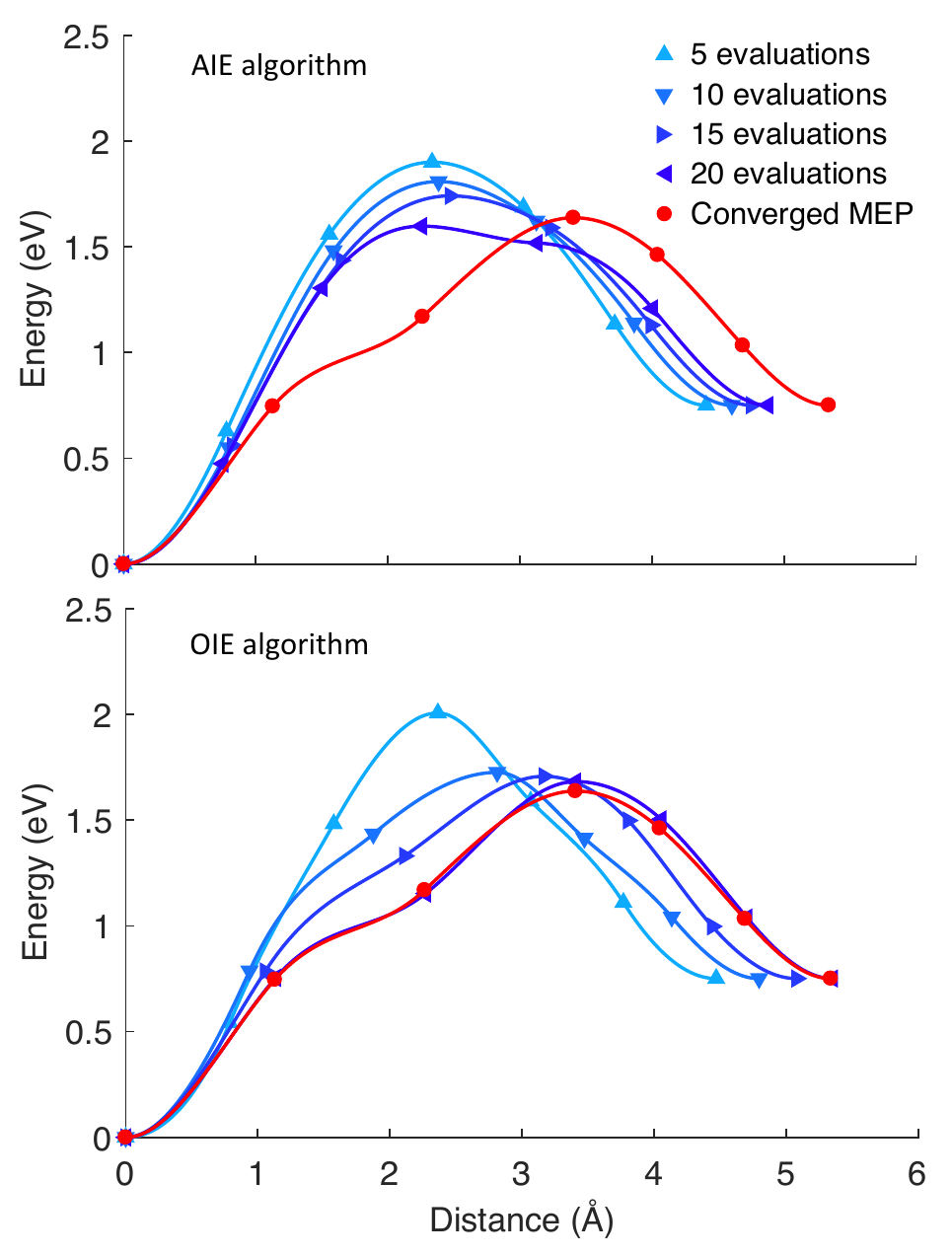}
\caption{
Comparison of the performance of the all-images-evaluated (AIE) and the one-image-evaluated (OIE) algorithms
(Hessian input used) for transition $F$ when the substrate atoms are frozen.
The markers show the GP-approximated energy at the images of the climbing image nudged elastic band relaxed
on the approximate energy surface after 5, 10, 15 and 20 energy and force evaluations (blue) and after final convergence (red).
With the AIE algorithm (upper figure), the variation of the energy is significantly different for the path
on the approximate surface compared with the true minimum energy path even after 20 evaluations (four Gaussian process regression iterations).
With the OIE algorithm (lower figure), the variation of the energy along the path on the
approximate surface is close to that of the converged MEP after 20 evaluations (corresponding to 20 GPR iterations in this case).
}
\label{fig:fig8}
\end{figure}

Even larger savings are obtained by using the uncertainty estimate provided by the GP model
to evaluate the energy and force only at the image where the true energy is the most poorly known
instead of all images, i.e., to move to the OIE from the AIE algorithm.
This is illustrated for transition $F$ in Fig.~\ref{fig:fig8}, where
the GP-approximated energy at the CI-NEB images
is shown after a certain number of energy and force evaluations for the AIE and OIE algorithms.
In the AIE case, the path is still qualitatively incorrect after 20 evaluations (four GPR iterations),
while this suffices for the OIE algorithm (20 GPR iterations) to nearly reach convergence.
For the AIE algorithm, it takes a total of 75 energy and force evaluations to confirm final convergence, while the OIE algorithm requires 39.
Similar reduction in the number of energy and force evaluations was found for the other transitions.

The number of energy and force evaluations required to confirm final convergence for each of
the 13 transitions using the AIE and OIE algorithms is given in Table~\ref{tab:table1}
as a fraction of the number of evaluations required by a regular CI-NEB method.
Also, the effect of using the Hessian input for the AIE algorithm is shown.
These numbers correspond to the case where six nearest substrate atoms can move in addition to the seven island atoms.
The relative number of evaluations compared to the regular CI-NEB varies between the transitions.
A clear trend is that the more complex the transition and the more iterations required by the regular CI-NEB,
the larger is the relative effect of the GPR approach.

\begin{table}[htbp]
\caption{\label{tab:table1}
The number of energy and force evaluations needed to converge the regular climbing image nudged elastic band (CI-NEB) calculations
of the heptamer island benchmark transitions (shown in Fig.~\ref{fig:fig5}) when 39 degrees of freedom are included, and the
reduction in the number of evaluations obtained with the Gaussian process regression approach
using the all-images-evaluated algorithm without the Hessian input (AIE),
all-images-evaluated algorithm with the Hessian input (AIE-H) and
one-image-evaluated algorithm without the Hessian input (OIE).
}
\vskip 0.4 true cm
\begin{center}
    \begin{tabular}{| c | c | >{\centering\arraybackslash}p{60pt} | >{\centering\arraybackslash}p{60pt} | >{\centering\arraybackslash}p{60pt} |}
    \hline
          & Number of   & \multicolumn{3}{c|}{Number of evaluations as a fraction of} \\
          & evaluations & \multicolumn{3}{c|}{evaluations needed for CI-NEB} \\ \hline
    Transition & CI-NEB & \multicolumn{1}{c|}{AIE} & \multicolumn{1}{c|}{AIE-H} & \multicolumn{1}{c|}{OIE} \\ \hline
    A          & 120    & 0.42                     & 0.42                       & 0.13 \\
    B          & 120    & 0.42                     & 0.42                       & 0.23 \\
    C          & 285    & 0.25                     & 0.21                       & 0.13 \\
    D          & 265    & 0.26                     & 0.23                       & 0.14 \\
    E          & 290    & 0.24                     & 0.19                       & 0.13 \\
    F          & 855    & 0.12                     & 0.12                       & 0.05 \\
    G          & 840    & 0.13                     & 0.11                       & 0.05 \\
    H          & 1480   & 0.08                     & 0.07                       & 0.04 \\
    I          & 1480   & 0.07                     & 0.07                       & 0.04 \\
    J          & 605    & 0.15                     & 0.12                       & 0.07 \\
    K          & 610    & 0.14                     & 0.12                       & 0.07 \\
    L          & 565    & 0.17                     & 0.12                       & 0.06 \\
    M          & 570    & 0.17                     & 0.11                       & 0.06 \\
    \hline
    \end{tabular}
\end{center}
\end{table}

\begin{figure}[htbp]
\centering
\includegraphics{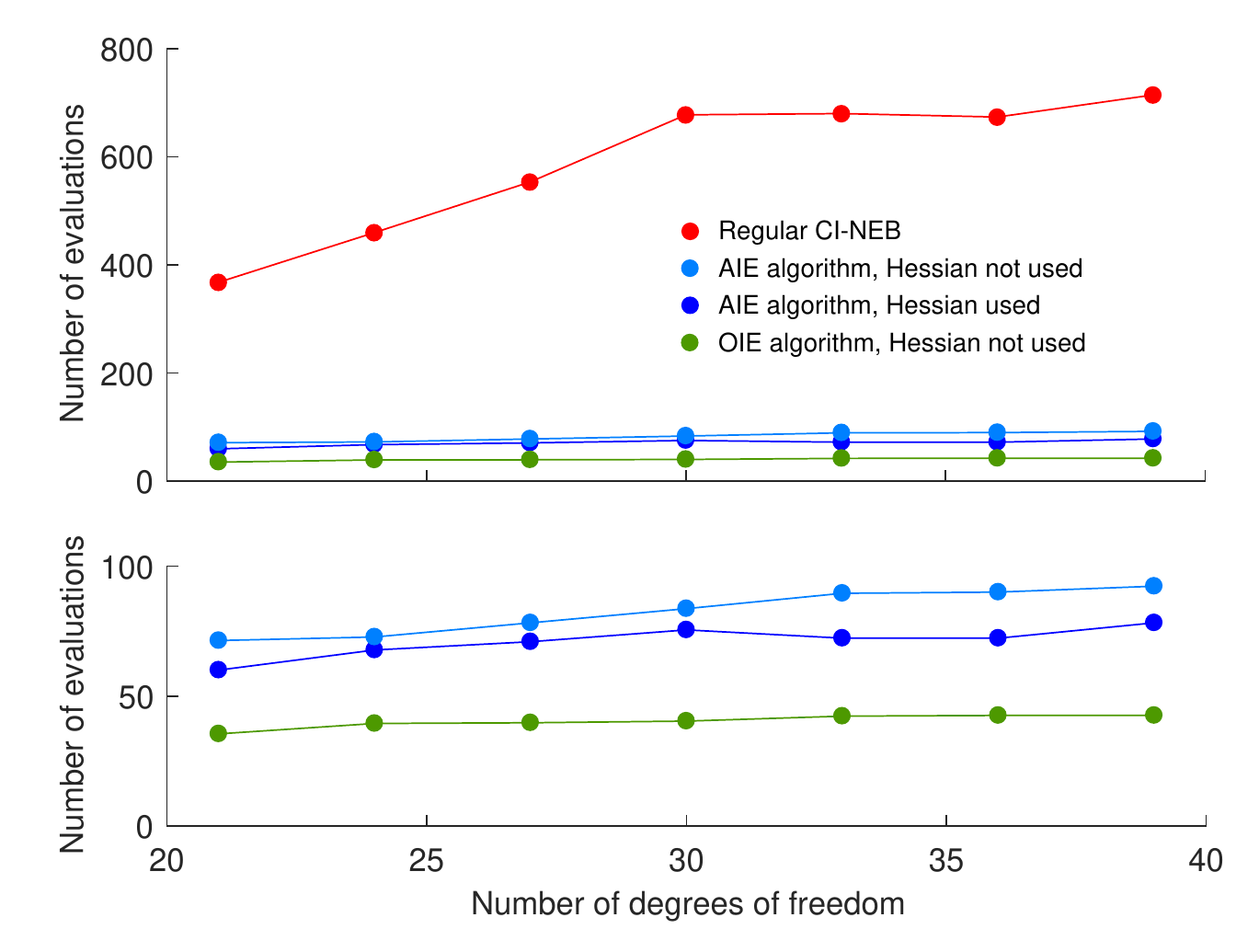}
\caption{
Average number of energy and force evaluations needed to converge to a minimum energy path in climbing image
nudged elastic band calculations of the heptamer island benchmark as a function of the number of
degrees of freedom, increased by allowing a larger number of substrate atoms to move.
The lower figure shows the same results as the upper figure on a different scale
to better distinguish between the various implementations of the Gaussian process regression approach.
For the larger numbers of degrees of freedom, the one-image-evaluated (OIE) algorithm provides about 1/20 reduction in the number of energy and force
evaluations as compared with a regular CI-NEB calculation.
The all-images-evaluated (AIE) algorithm requires about twice as
many evaluations as the OIE algorithm, but the use of the Hessian at the initial and
final state minima can reduce that by about 20\%. The use of the Hessian has less effect when the
OIE algorithm is used (not shown).
}
\label{fig:fig9}
\end{figure}

The average number of energy and force evaluations for transitions $C$ -- $M$ as a function of the number of degrees of
freedom is shown in Fig.~\ref{fig:fig9}.\footnote{Transitions $A$ and $B$ are not included in the averages shown in Fig.~\ref{fig:fig9} because the regular CI-NEB required
an anomalously large number of iterations for some of the intermediate numbers of degrees of freedom.
The results of the GPR algorithms were, however, similar for all numbers of degrees of freedom tested here.}
For the smallest number of degrees of freedom, 21, only the seven island atoms are allowed to move while all the 
substrate atoms are frozen. For the larger numbers of degrees of freedom, some of the surface atoms are also allowed to move.
Starting with the AIE algorithm, the use of the Hessian input reduces the number of evaluations by about 20\%,
but the transition to the OIE algorithm has an even larger effect, a reduction to a half.

The OIE results represent savings of an order of magnitude with respect to the regular CI-NEB calculation.
The number of energy and force evaluations reported here for the CI-NEB method is similar to what has been reported
earlier for this test problem \cite{Henkelman00,Chill2014}.
It is possible to use a more efficient minimization scheme
to relax the images in CI-NEB calculations \cite{Sheppard08}, but the difference is not so large.
The test results presented here, therefore, show that the use of GPR can significantly reduce the
computational effort in, for example, calculations of MEPs for surface processes.


\section{Discussion}
\label{sec:Discussion}

The results presented here show that the GPR approach can reduce the
number of energy and force evaluations needed in CI-NEB calculations of MEPs by an order of magnitude.
This is important since a large amount of computer time is
used in such calculations, especially when {\it ab initio} or density functional theory calculations are used.
Compared with the previous proof-of-principle calculations \cite{Koistinen16},
three major improvements to the algorithm have been presented here.
First of all, the CI-NEB algorithm was used where one of the images is pushed up to the maximum along the MEP.
This provides stability and accelerates convergence because it focuses more evaluations of the energy
and force in the vicinity of the first-order saddle point, the most important part of the energy surface.

Secondly, the benefit of using the finite difference estimate of the Hessian at the endpoints was demonstrated and found to 
result in a 20\% reduction in the number of energy and function evaluations needed to converge on an MEP
with the AIE algorithm. This estimation of the Hessian does not represent any additional energy and force evaluations
in cases where the goal is to calculate the transition rates using harmonic transition state theory.
The usual ordering of the calculations is simply changed; the Hessian at the endpoint minima is evaluated before the MEP
calculation rather than afterwards.
If a higher level of rate theory, such as optimal hyperplanar transition state theory \cite{Johannesson00}, is used, then 
analogous information about the initial and final states can be obtained from dynamical trajectories.

While the Hessian input reduced the number of energy and function evaluations only by 20\%, a significant advantage is likely in greater
stability and lower probability of the path escaping into irrelevant regions of the energy surface in the first few GPR iterations.
This we have seen in preliminary studies of dissociative adsorption of H$_2$ on a metal surface and water molecule
diffusion on an ice Ih(0001) surface \cite{Batista01}. The addition of the finite difference data points for the Hessian adds
significantly to the memory requirements of the GPR calculation. 
Those data points could, however, be dropped after just a few GPR iterations since they are not needed when the GP 
approximation of the energy surface has become reasonably accurate around the MEP.
A more elegant and efficient way of incorporating the information from the Hessian into the covariance calculation could 
also be developed. The implementation described here represents just an initial test to see how important such information can be.

As a third improvement, a significant reduction in the computational effort was shown to be possible by using the one-image-evaluated algorithm
instead of the earlier all-images-evaluated approach. The number of energy and force evaluations was reduced to a half in the heptamer benchmark.
In the OIE algorithm, the true energy and force are evaluated only at one image, rather than all images along the path, before a new GPR iteration.
A calculated choice of the location of each evaluation can be made based on the uncertainty estimate provided by the GP model.
Interestingly, the use of the Hessian did not provide significant reduction in the number
of energy and force evaluations required by the OIE algorithm.
This apparently stems from the fact that the OIE algorithm involves fewer evaluations in the early
phase of the iterative GPR process where the approximation to the energy surface is poor.

The heptamer island benchmark studied here is a relatively simple example, and it will be important to test the GPR approach on more complex systems to
be able to fully assess its utility and to develop the methodology further.
On complex energy surfaces, there may exist multiple MEPs connecting the two endpoint minima,
which would require some kind of sampling of MEPs \cite{Maras16}.
Also, some systems may have multiple local minima and highly curved MEPs,
which can lead to convergence problems unless a large number of images are included in the calculation.
In systems where various types of molecular interactions are involved,
the optimal length scale may vary depending on the location in the coordinate space.
In such cases, it may be advantageous to use a GP model that allows different
length scales in different parts of the space.

In order to tackle large systems, the scaling of the GPR calculations will need to be improved.
A more efficient implementation could be obtained, e.g., by using a compactly supported covariance function
to produce a sparse covariance matrix where data points far away from each other become independent \cite{Vanhatalo10}.
It may also be possible to reduce the dimensionality by using partially additive models,
where the interaction term in the energy function for far away atoms is ignored.
There is, however, a large set of important problems, such as calculations of catalytic processes often involving rather small molecules adsorbed on
surfaces, where the complexity is comparable to the heptamer island benchmark and where the GPR approach is
clearly going to offer a significant reduction in computational effort in NEB calculations of MEPs.

At a low enough temperature, quantum mechanical tunneling becomes the dominant transition mechanism,
and the task is then to find a minimal action path \cite{Jonsson11,Mills97,Mills98}.
The effect of the GPR approach in tunneling path calculations could be even larger than for MEP calculations,
since each iteration involves more energy and force calculations (Feynman paths
rather than points in configuration space) and thereby more data for the modeling.
In addition to atomic rearrangements, it will be valuable to apply the GPR approach also 
to magnetic transitions where the magnetic properties of the system are evaluated by computationally intensive 
{\it ab initio} or density functional theory calculations. 
The relevant degrees of freedom in magnetic transitions are the angles defining the orientation of
the magnetic vectors, and the task is again to find MEPs on the energy surface with respect to those angles \cite{bessarab_13a,bessarab_14,bessarab_14b}.


\section*{Acknowledgements}

We thank Professor Andrew Peterson for helpful discussions.
This work was supported by the Icelandic Research Fund and the Academy of Finland (Grant No. 278260).
We acknowledge the computational resources provided by the Aalto Science-IT project and the Computer Services of the University of Iceland.



\newpage

\end{document}